\documentclass[prl,twocolumn,amssymb,showpacs]{revtex4}

\usepackage{graphicx}
\usepackage{amssymb,amsmath}

\begin{document}

\title{Quantum  fidelity in the thermodynamic limit}

\author{Marek M. Rams$^{1,2}$ and Bogdan Damski$^1$} 
\affiliation{$^1$Los Alamos National Laboratory, 
Theoretical Division, MS B213, 
Los Alamos, NM, 87545, USA \\
$^2$Institute of Physics, Jagiellonian University, Reymonta 4, PL-30059 Krak\'ow, Poland} 
\begin{abstract}
We study quantum fidelity, the overlap between two ground states of a many-body system,
focusing on the thermodynamic regime. We show how drop of fidelity near a critical point 
encodes universal information about a quantum phase transition. Our general scaling results 
are illustrated in the quantum  Ising chain for which 
a remarkably simple  expression  for fidelity is found.
\end{abstract}
\maketitle

A quantum phase transition (QPT) happens when dramatic changes in the 
ground state properties of a quantum system can be induced by 
a tiny  variation of an external parameter \cite{sachdev}. 
This external parameter
can be a strength of a magnetic field in spin systems 
(e.g. Ising chains \cite{Coldea2010etal} and spin-1 Bose-Einstein condensates 
\cite{sadler2006etal}),
intensity of a laser beam creating a lattice for cold atom 
emulators of Hubbard models \cite{Hubbard_exp}, or 
dopant concentration in high-Tc superconductors \cite{Lee2006}.
At the heart of the sharp transition lies
non-analyticity of the ground state wave-function across the critical 
point separating the two phases.  
QPTs, traditionally associated with condensed matter physics, 
are nowadays intensively studied from the quantum information perspective 
 (see e.g. \cite{Osterloh2002}).   

Quantum fidelity -- also referred to as fidelity -- is 
a popular concept of quantum information science defined here as 
the overlap between two quantum states
\begin{equation}
{\cal F}(g,\delta) = |\langle g-\delta|g+\delta\rangle|,
\label{def_F}
\end{equation}
where $|g\rangle$ is a ground state wave-function of a many-body Hamiltonian
$\hat H(g)$ describing the system exposed to an external field $g$, and
$\delta$ is  a small parameter difference.
It provides the 
most basic probe into the dramatic change of the wave-function 
near and at  the  critical point \cite{Zanardi2006}.

The recent surge in studies of fidelity follows 
discovery that quantum 
criticality promotes  decay of  fidelity \cite{Zanardi2006}.
This is in agreement with the intuitive picture of a QPT: 
as system properties change dramatically in the neighborhood
of the critical point, ground state wave-function taken at  
different values of the external parameter --
$|g-\delta\rangle$ and $|g+\delta\rangle$ -- have little in common
and so their overlap  decreases. 

As  fidelity is given by the angle between two vectors 
in the Hilbert space, 
it is a geometric quantity \cite{Zanardi_geometric2}. Thus, it has been proposed 
as a robust geometric probe of quantum criticality 
applicable to all  systems undergoing a QPT {\it regardless}
of their symmetries and order parameters whose prior knowledge is 
required in traditional  approaches to QPTs. 
Fidelity has been recently studied in this context in several models of condensed matter 
physics (see \cite{Gu2008} and references therein).

Besides being an efficient probe of quantum criticality,  fidelity 
appears in numerous problems in quantum physics.
Indeed, it is related to density of topological defects after 
a quench   \cite{BDPRL2005,barankov,polkovnikov}, decoherence rate 
of a test qubit interacting with an out-of-equilibrium environment \cite{BD2009},
orthogonality catastrophe of condensed matter systems (see \cite{Anderson1967} 
and the references citing it).
Therefore its understanding has an interdisciplinary impact.
 
To unravel the influence of quantum criticality on fidelity, one has to determine if its 
drop  near the critical point  encodes universal information about the transition
in addition to providing the location of the critical point. This universal information is
given by the critical exponents and reflects symmetries of the model rather than its microscopic details. 
In the ``small system limit'', which broadly speaking corresponds to $\delta\to0$ at fixed system size $N$, the answer is positive. 
This is explored in the fidelity susceptibility approach
where    \cite{Zanardi2006,You2007,Gu2008}
\begin{equation}
{\cal F}(g,\delta) \simeq 1 - \delta^2\chi_F(g)/2,
\label{def_sus}
\end{equation}
and $\chi_F$ stands for fidelity susceptibility. Universal information, or simply the critical exponent $\nu$,
 is encoded in its scaling: at the critical point 
 $\chi_F(g_c) \sim N^{2/d \nu}$, while far away from it $\chi_F(g) \sim N|g-g_c|^{d\nu-2}$, where $d$ is system dimensionality
   \cite{ABQ2010,barankov,polkovnikov}.
 
 In the  thermodynamic limit, which broadly speaking corresponds to $N\to\infty$ at fixed $\delta$, the answer is positive as well.
 This is our key result stating that 
 \begin{equation}
\ln{\cal F}(g,\delta) \simeq  -N|\delta|^{d\nu} A\left(\frac{g-g_c}{|\delta|} \right),
\label{near_F}
\end{equation}
where $A$ is a scaling function. In particular, we see that fidelity is non-analytic in $\delta$ at the critical point, 
$\ln{\cal F}(g_c,\delta) \sim -N|\delta|^{d\nu}$, while away from it, i.e., for 
$|\delta| \ll |g-g_c| \ll 1$, we obtain
\begin{equation}
\ln{\cal F}(g,\delta) \sim -N\delta^2 |g-g_c|^{d\nu -2},
\label{away_fid}
\end{equation}
after expansion of the scaling function.
These results, in particular,  set firm foundations for usage of fidelity as a probe of quantum 
criticality in thermodynamically-large systems. In the context of theoretical studies of QPTs, 
the strength of the fidelity approach 
lies in its simplicity: all  information encoded in the ground state wave-function(s)   
is ``compactified'' into a single number. A competing approach
for extraction of  the exponent $\nu$ -- study of the asymptotic decay of 
correlation functions to obtain the correlation length -- is considerably more complicated.
Below we  illustrate these predictions on the paradigmatic model of quantum phase
transitions, the Ising chain, 
and discuss the scaling theory that leads to (\ref{near_F}) and (\ref{away_fid}).

The Hamiltonian of the one dimensional Ising chain reads \cite{sachdev}
$$
\hat H(g) = 
-\sum_{i=1}^N(\sigma^x_i\sigma^x_{i+1} +  g \sigma^z_i),
$$
where $g$ stands for a magnetic field acting along the $z$ direction. 
Above the spin-spin interactions try to enforce $\pm x$ polarization 
of spins, while the magnetic field tries to polarize spins 
along its direction ($+z$ for $g>0$).
This competition results in two critical points  at $g_c=\pm1$: the 
system is in the ferromagnetic (paramagnetic) phase for $-1<g<1$ ($|g|>1$).
The critical exponent $\nu=1$.
This model is solved  by mapping spins onto 
non-interacting fermions via the Jordan-Wigner transformation  \cite{sachdev}.

\begin{figure}
\includegraphics[width=\columnwidth,clip=true]{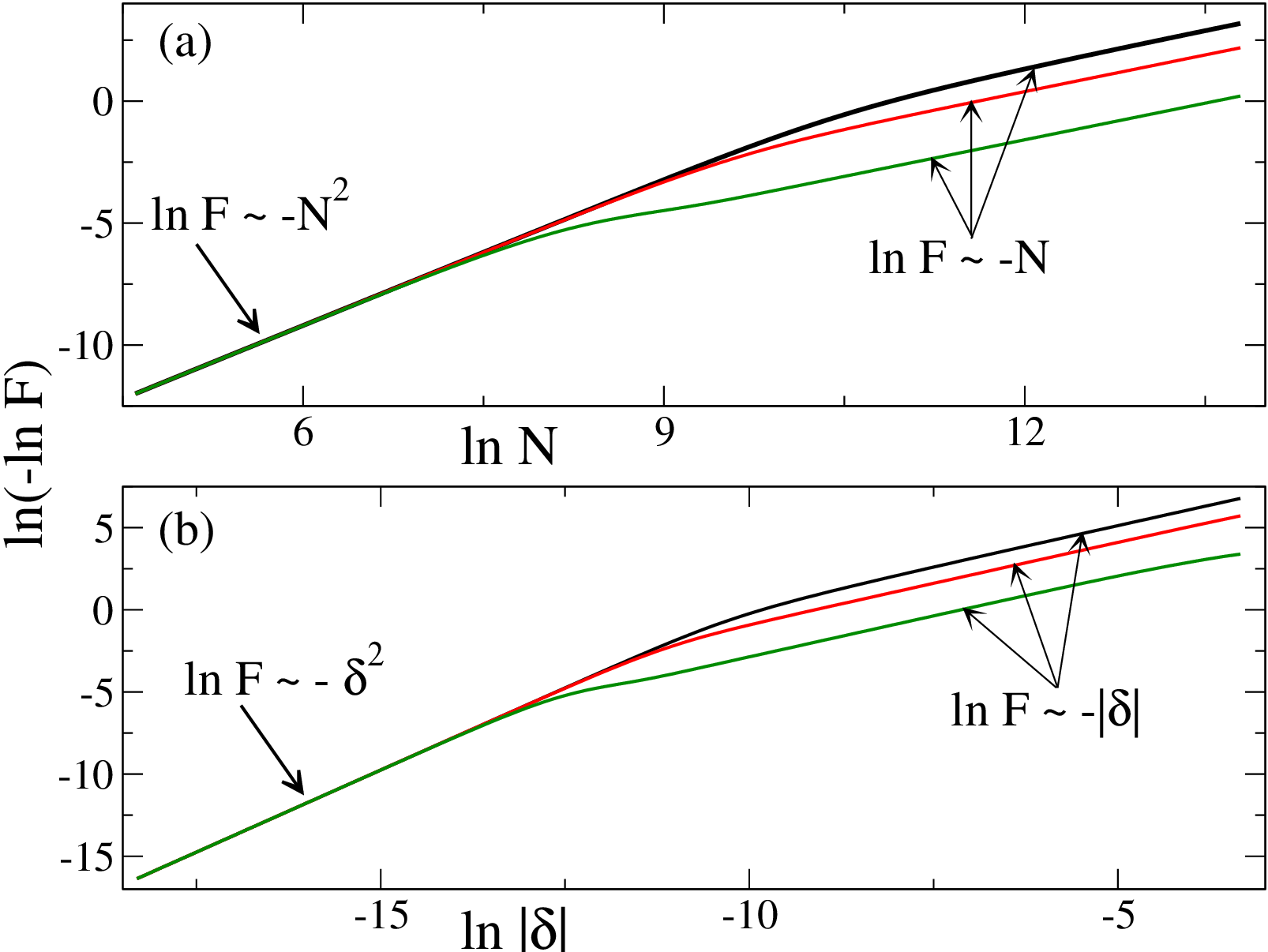}
\caption{(color online) Fidelity of the Ising chain 
near the critical point  as a function of (a) the system size $N$ at 
fixed $\delta=10^{-4}$ and  (b) parameter difference $\delta$ at the fixed
system size $N=10^5$.
On both panels the curves from top to bottom correspond to 
${\cal F}(1,\delta)$, ${\cal F}(1+\delta, \delta)$ and ${\cal F}(1+5\delta,\delta)$. 
}
\label{fig1}
\end{figure}

\begin{figure}
\includegraphics[width=\columnwidth,clip=true]{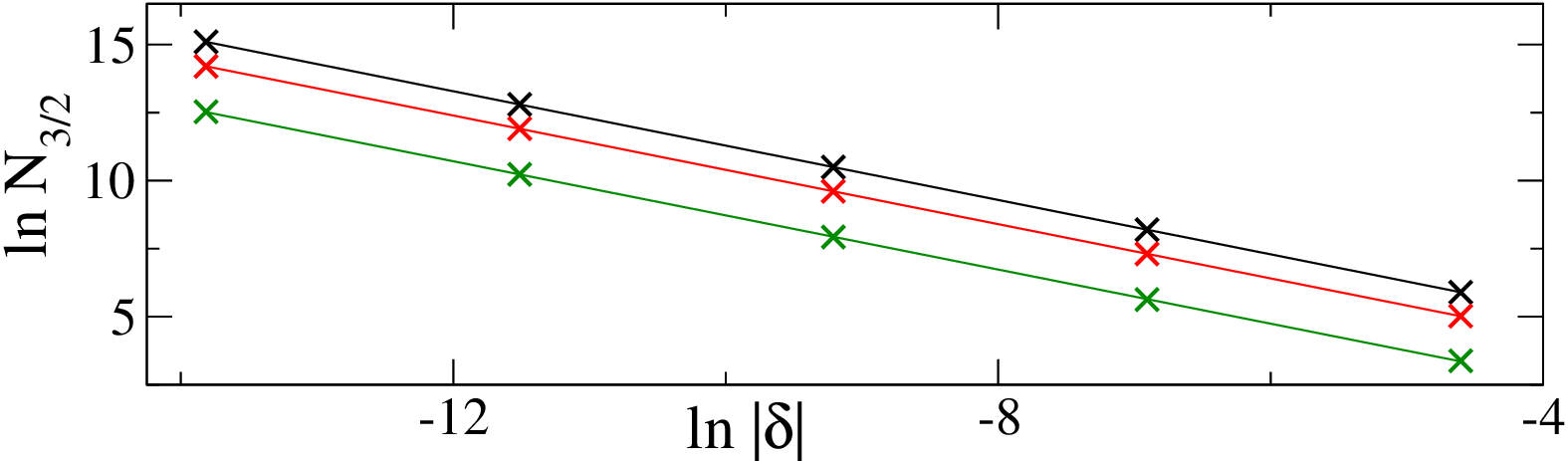}
\caption{(color online) Study of the crossover between the 
``small system limit'' and the thermodynamic limit  illustrated  in Fig. 
\ref{fig1}. 
As the system size is increased in Fig. \ref{fig1}a,  the slope of the curves 
changes smoothly from $2$ (corresponding to $\ln{\cal F}\sim -N^2$) to $1$
(corresponding to $\ln{\cal F}\sim -N$). The crossover region
between the two limits is centered around $N=N_{3/2}$ where the slope equals $3/2$.
To find it, we have calculated numerically
${\cal F}(g,\delta)$ vs. $N$ -- as in Fig. \ref{fig1}a -- for various $\delta$'s and found that 
$N_{3/2}|\delta|\sim1$.  
This is illustrated 
in this figure where data sets from top to bottom correspond to results obtained for $g=1$, $1+\delta$ and $1+5\delta$,
respectively (similarly as in Fig. \ref{fig1}a). 
The power-law fits (straight lines) to numerical data (crosses) 
give $N_{3/2}=a|\delta|^{-b}$, where $b= 0.995\pm0.003$ for all
three fits, while the prefactor $a$ changes between the fits from $3.6$ to $0.3$.
Similar  analysis can be done on curves shown in Fig. \ref{fig1}b
providing the same result. Thus we conclude that the crossover condition reads
$N|\delta|\sim 1$ near the Ising critical point.
}
\label{fig2}
\end{figure}

Behavior of fidelity (\ref{def_F}) around the critical 
point, $g\approx g_c$, is summarized in Fig. \ref{fig1}.
In Fig. \ref{fig1}a the parameter difference $\delta$ is 
kept fixed and the system size is increased. For small system
sizes we reproduce the known result, $\ln{\cal F} \sim -N^2$ \cite{Zanardi2006}, 
resulting from finite size 
scaling effects (see e.g. \cite{ABQ2010,barankov,polkovnikov,Gu2008}).
For large system sizes, however, we obtain 
$\ln{\cal F}\sim -N$ in  agreement with (\ref{near_F}) and  the fidelity 
per site approach  \cite{Zhou2008,Zhou2008_Vidal,Zhou2008_Ising}. 
As is shown in Fig. \ref{fig2}, 
the transition between the two regimes takes place when 
\begin{equation}
N|\delta|\sim 1,
\label{Ndelta}
\end{equation}
which will be discussed below.

Similarly, we observe two distinct regimes when 
the system size $N$ is kept fixed and the parameter difference $\delta$
is varied (Fig. \ref{fig1}b). For $N|\delta|\ll1$ we observe 
$\ln{\cal F}\sim-\delta^2$, in agreement with (\ref{def_sus}), 
while for $N|\delta|\gg1$ we find $\ln{\cal F}\sim -|\delta|$ in  agreement
with (\ref{near_F}).
In the latter fidelity {\it approaches} non-analytic limit
(where $\partial_\delta {\cal F}$ at $\delta=0$ is undefined)
reflecting singularities of the ground state wave-function 
resulting from the QPT
 \cite{note}.  

We also see on both panels of Fig. \ref{fig1} that
all curves collapse for $N|\delta|\ll1$, while they 
stay distinct in the opposite limit. Thus, for  $N|\delta|\gg1$
sensitivity of fidelity to quantum criticality 
is enhanced. This can be  understood if we focus on Fig. \ref{fig1}a:
in the large $N$ limit  dramatic changes in the 
ground state wave-function near the critical point  are expected.

\begin{figure}
\includegraphics[width=\columnwidth,clip=true]{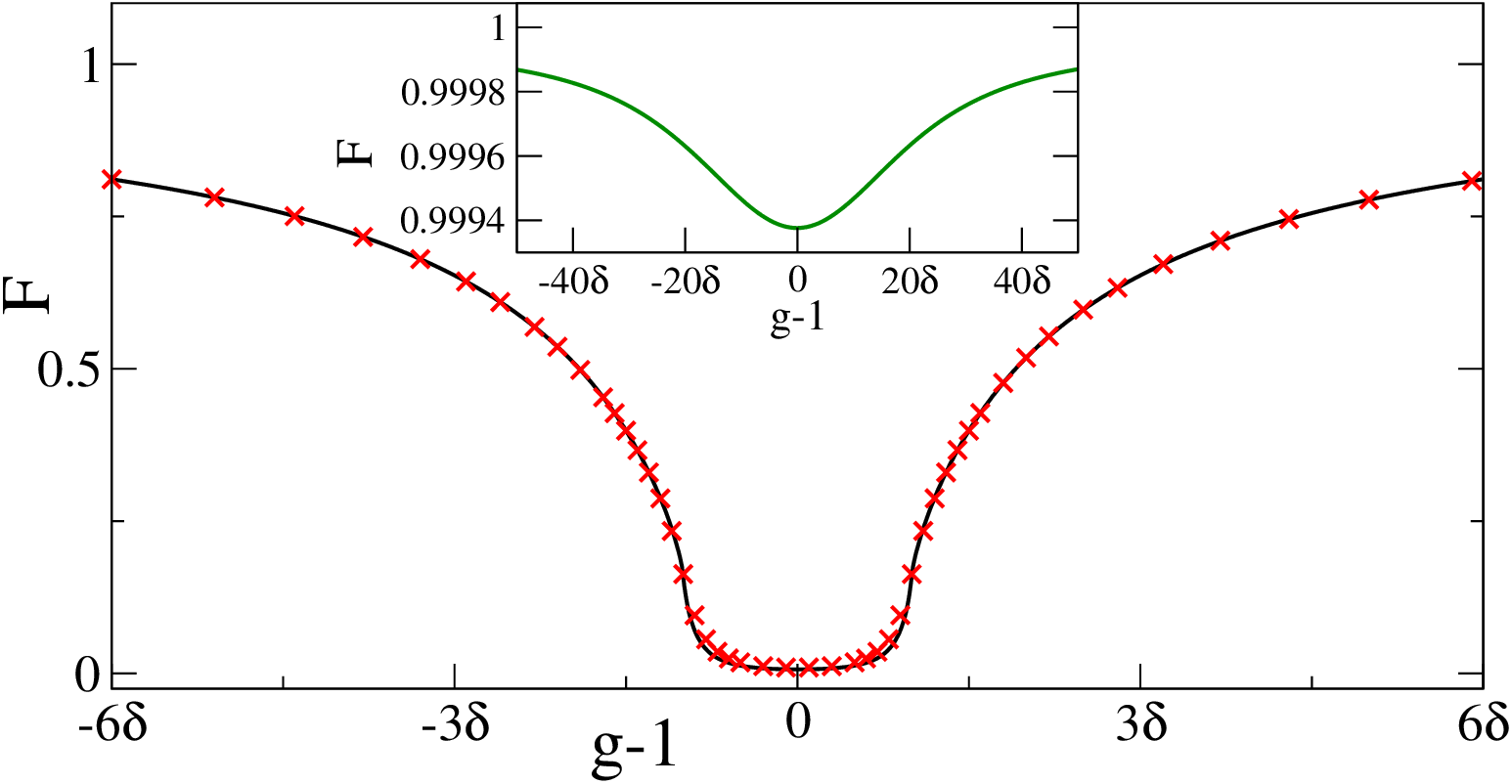}
\caption{(color online) Fidelity ${\cal F}(g,\delta)$ of the Ising chain 
near the critical point: thermodynamic limit (main plot) vs. ``small system limit'' (inset).
Main plot: black curve is our analytic approximation (\ref{fid_ising}), while 
red crosses come from numerics. Both were obtained for $N= 2\times 10^5$ and 
$\delta=10^{-4}$ ($N|\delta|\gg1$).
Inset: numerical result for $N=10^3$ and $\delta=10^{-4}$ ($N|\delta|\ll1$).
In the ``small system limit'' fidelity stays close to unity at any distance 
from the critical point, while in the thermodynamic limit it can interpolate
between zero and unity.
}
\label{fig3}
\end{figure}

\begin{figure}
\includegraphics[width=\columnwidth,clip=true]{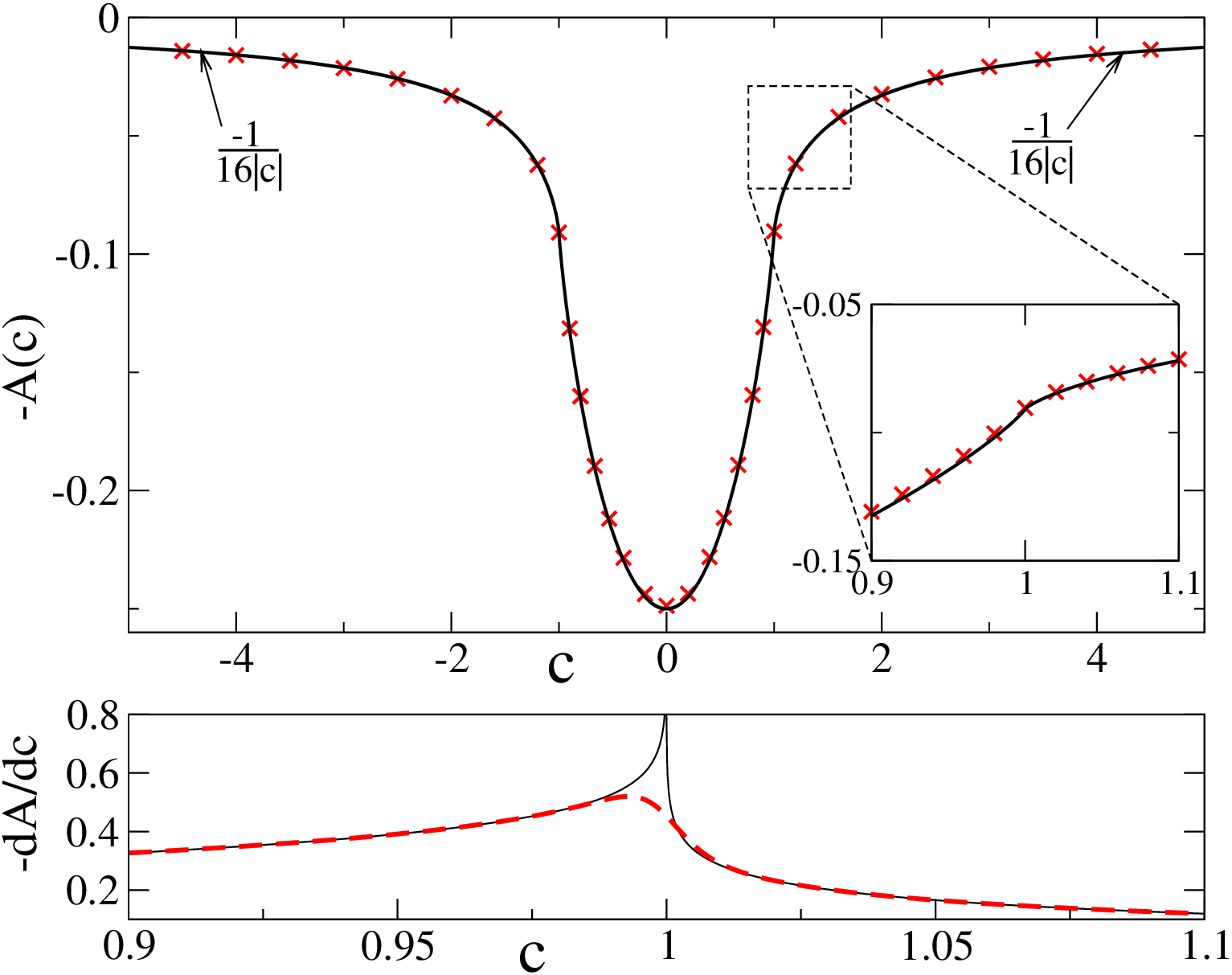}
\caption{(color online) 
Upper plot: scaling function $A(c)$ of the Ising chain. 
The solid black line provides the analytic result (\ref{Ac}), while the 
red crosses show numerics (i.e., $\ln {\cal F} / N |\delta| $). 
The inset highlights singularity at  $c=1$.
Lower plot: logarithmic divergence of $dA/dc|_{c=1}$
discussed in the text. The solid black line 
is the derivative of (\ref{Ac}), while the red dashed line 
is a numerical result: the difference between the two near the pinch point at $c=1$ 
is due to the finite system 
size $N$ \cite{Zhou2008_Ising}. It disappears in the limit of  $N\to\infty$. In both plots numerics is done for 
$N=10^5$ and $\delta=\pi 10^{-3}$.
}
\label{fig4}
\end{figure}

As analytical results for  fidelity are
scarce, we find it remarkable that we can derive 
 accurate analytical description  in the complicated 
limit of $N|\delta|\gg1$,
where the  Taylor expansion (\ref{def_sus}) fails.
To proceed, we calculate ${\cal F}(1+\epsilon,\delta)$,
where $\epsilon$ measures distance from the critical 
point. For the Ising chain 
${\cal F}~=~ \Pi_{k>0}f_k$, where 
$f_k=\cos(\theta_+(k)/2-\theta_-(k)/2)$ and
$\tan(\theta_\pm(k)) = \sin k/(1+\epsilon\pm\delta-\cos k)$. 
We stay close to the critical point so that 
$0\le|\delta|,|\epsilon|\ll 1$ and introduce natural parameterization: 
 $c=\epsilon/|\delta|$.
Taking the limit of $N\to\infty$ at {\it fixed} $\delta$ the 
product $\Pi_k f_k$ can be changed into 
$\exp(N\int dk \ln f_k/2\pi)$, which can be further simplified to 
\begin{equation}
\ln{\cal F} ~\simeq~ -N |\delta| A(c)
\label{fid_ising}
\end{equation}
in the leading order in $\delta$ and $\epsilon$. This result is in prefect agreement with our universal scaling law 
(\ref{near_F}): note that $\nu,d=1$ in our model and $c=(g-g_c)/|\delta|$.
Moreover, it agrees well with exact numerical simulations: Fig. \ref{fig3}.
Above $A(c)$  is given by
\begin{equation}
A(c)= \left \{
\begin{array}{c} 
\begin{split}
& \frac{1}{4}+ \frac{|c| K(c_1)}{2\pi}+\frac{(|c|-1) {\rm Im} E(c_2)}{4\pi} ; |c|\le1  \\
& \frac{|c|}{4} - \frac{|c| K(c_1)}{2\pi}-\frac{(|c|-1) {\rm Im} E(c_2)}{4\pi} ; |c|>1.
\end{split}
\end{array}
\right. 
\label{Ac}
\end{equation}
where $c_1 = -4 |c|/(|c|-1)^2$, $c_2=(|c|+1)^2/(|c|-1)^2$, and 
$K$ and $E$ are complete elliptic integrals of the first and second kind, respectively.
Agreement between (\ref{Ac})
and numerics is very good: see Fig. \ref{fig4} for detailed comparison of 
$A(c)$ to numerics. 
Several interesting results can be obtained now.

First, Eq. (\ref{fid_ising}) shows analytically how the so-called Anderson 
catastrophe -- disappearance of the overlap between distinct ground states
of an infinitely large many-body quantum system 
\cite{Anderson1967}
-- happens in the Ising chain.

Second, Eq. (\ref{fid_ising}) explains 
the lack of collapse of the various curves 
providing fidelity around the critical point
in the $N|\delta|\gg1$ limit. Indeed, 
fidelity  calculated for two ground states symmetrically
around the critical point is  ${\cal F}(1,\delta)\simeq\exp(-N|\delta|/4)$,
but if one of the ground states is obtained  
at the critical point,  
${\cal F}(1\pm\delta,\delta) \simeq \exp(-N|\delta|(\pi-2)/4\pi)$. 
In the opposite limit of $N|\delta|\ll1$, 
${\cal F}\simeq 1-\delta^2 N^2/16$ in both cases explaining  the collapse of
all curves in this limit in Fig. \ref{fig1}. 

Third, there is a singularity in the derivative of fidelity 
when one of the states is calculated at the critical point:
$d{\cal F}(g\pm\delta,\delta)/dg|_{g=g_c=1}$ is divergent
when $N\to\infty$ at fixed $\delta$.
This reflects singularity of the wave-function at the critical point 
approached in the thermodynamic limit. Quantitatively, 
$dA(c)/dc|_{c\to1^{\pm}}= \ln|1-c|/4\pi - 3\ln2/4\pi+(1\pm1)/8 + 
{\cal O}((1-c)\ln|1-c|)$, which is logarithmically divergent at 
$c=1$ (Fig. \ref{fig4}).
This divergence is a signature of a pinch point found in
\cite{Zhou2008,Zhou2008_Vidal,Zhou2008_Ising}  
when fidelity between two distinct ground states states
 was studied. 
The logarithmic divergence in the Ising chain
was numerically observed in 
\cite{Zhou2008_Ising}.

Last but not least, we obtain from (\ref{fid_ising})
a compact expression for fidelity away from the critical 
point. Taking  $|c| \gg 1$ (but still $|\epsilon| = |c\delta| \ll 1$), 
$A(c)\simeq 1/16 |c|$ and so 
\begin{equation}
{\cal  F}~\simeq~\exp(-N\delta^2/16|\epsilon|),
\label{f_away}
\end{equation}
in agreement with (\ref{away_fid}).
This reduces to a known result for fidelity susceptibility 
when the argument of the exponent is small and so 
${\cal F}\simeq 1-\delta^2N/16|\epsilon|$ 
(see e.g.  \cite{Gu2008}), but provides a new  
result in the opposite limit where lowest order
of the Taylor expansion is insufficient. We notice also 
that (\ref{f_away}) is analytical in $\delta$ even in the limit
of $N\to\infty$: 
there are 
no singularities expected when the system is far away from the critical 
point.

Below we derive general scaling results (\ref{near_F}) and (\ref{away_fid}).
This can be done by  studying  
the  scaling parameter
$$
\tilde d(g+\delta,g-\delta) = -\lim_{N\rightarrow \infty } \ln{\cal F}(g,\delta)/N, 
$$
introduced in  \cite{Zhou2008} in the context of fidelity per site
approach to the thermodynamic limit.
We expect that this limit is reached when 
\begin{equation}
\min[\xi(g+\delta), \xi(g-\delta)] \ll L,
\label{tlimit}
\end{equation}
where $\xi(g)$ is the correlation length at magnetic field $g$ and
$L$ is the  linear size of the system ($N=L^d$ for a  $d$-dimensional
system). Indeed, the smaller of the two correlation lengths
sets the  scale on which the states entering  fidelity 
 ``monitor'' each other (\ref{def_F}). In particular, it explains
our results showing that the thermodynamic limit is reached 
even when one of the states is calculated at the critical point 
and so its correlation length is infinite. Near a critical point 
 (\ref{tlimit}) is equivalent to $L|\delta|^\nu\gg1$ \cite{remark}.
 For the Ising chain studied above it reads $N|\delta|\gg1$ properly 
predicting the crossover condition  (\ref{Ndelta}) obtained 
from numerical simulations (Fig. \ref{fig2}).

Generalizing the scaling theory of  second order
QPTs (Sec. 1.4 of  \cite{ContinentinoBook}), we propose the following 
scaling ansatz for the universal part of the scaling parameter
$$
\tilde d(g_c+\epsilon+\delta,g_c+\epsilon-\delta)=b^{-d} f((\epsilon+\delta)b^{1/\nu},(\epsilon-\delta)b^{1/\nu}),
$$
where $f$ is the scaling function, $b$ is the scaling factor, and $\nu$ is 
the critical exponent providing divergence of the coherence length
$\xi\sim |g-g_c|^{-\nu}$. 
The scaling function  depends on both $\epsilon+\delta$ and $\epsilon-\delta$ 
as they are renormalized simultaneously. 
The factor $b^{-d}$ appears for dimensional reasons.
Scaling of $\epsilon+\delta$ and $\epsilon-\delta$ is given by scaling of the correlation length
$\xi(\epsilon\pm\delta)=b \xi((\epsilon\pm\delta)b^{1/\nu}) $.

Taking $g=g_c+\epsilon$, introducing natural  parameterization $\epsilon = c |\delta|$, and 
fixing the scale of renormalization through $|\delta|b^{1/\nu}=1$  we obtain
$\tilde d(g+\delta,g-\delta)=|\delta|^{d\nu} f(c+1,c-1)$. It  gives  
(\ref{near_F}) after setting $f(c+1,c-1)=A(c)$. In a general context,  (\ref{near_F})
shows how universal part of the scaling parameter causes 
the  Anderson catastrophe near a critical point.  

The scaling function $A(c)$ can be simplified away from the critical point.
We assume below $\epsilon,\delta>0$ for simplicity,  take
$\delta\ll \epsilon \ll 1$, and 
set $b$ through $(\epsilon+\delta)b^{1/\nu}=1$ exploring the freedom to choose the renormalization scale.
Simple calculation results in 
$\tilde d(g+\delta,g-\delta)=(\epsilon+\delta)^{d \nu} f(1,(\epsilon-\delta)/(\epsilon+\delta))$,
where the second argument of $f$ is close to unity. Expanding 
$f$ in it we get 
$\tilde d(g+\delta,g-\delta)\approx 2 \delta^2 \epsilon^{d \nu -2}  f''(1,x)|_{x=1}$ as
$f(1,x)$ has a minimum equal to zero at $x=1$. Thus, away from a critical point 
we end up with (\ref{away_fid}).
When the system is small enough, $N\delta^2 |\epsilon|^{d \nu -2}\ll 1$,
but still in the thermodynamic limit (\ref{tlimit}), we reproduce the known result
for fidelity susceptibility
$1-{\cal F} \sim  \delta^2 N |\epsilon|^{d \nu -2}$
\cite{barankov,polkovnikov,ABQ2010}. 
Otherwise,  (\ref{away_fid}) provides a new result.

On general grounds, one can expect that 
for systems with $d\nu\ge2$ non-universal (system-specific) corrections 
to the above scaling relations may be significant \cite{polkovnikov}, which requires further investigation.

Summarizing, our work characterizes  fidelity -- a modern probe of quantum criticality --
in the thermodynamic limit.
We have derived, and verified on a specific model, new universal scaling properties 
of fidelity. These findings should be experimentally relevant as the first 
experimental studies of  ground state fidelity have been
already done \cite{Zhang_all}.

This work is supported by U.S. Department of Energy through the LANL/LDRD Program.

\end{document}